Research Article (Research Article)

# Improving Financial Forecasting with a Synergistic LLM-Transformer Architecture: A Hybrid Approach to Stock Price Prediction

**Sayed Akif Hussain[1*], Chen Qiu-shi[1], Syed Amer Hussain[2], Syed Atif Hussain[3]**

**Asma Komal[4], Muhammad Imran Khalid[4],**

[1] Management Science & Engineering, CQUPT, Chongqing, 400065, China
[2] Electrical & Electronics Engineering, COMSATS Abbottabad Campus, Abbottabad, 22020, Pakistan
[3] NUST, Collage of EME, Rawalpindi, Pakistan
[4] Computer Science & Technology, CQUPT, Chongqing, 400065, China
* Corresponding author: Author A (syedakifhussain110@gmail.com)
Author B (chenqiushi@cqupt.edu.cn )

*Abstract*: This study proposes a novel hybrid deep learning framework that integrates a Large Language Model (LLM) with a Transformer architecture for stock price forecasting. The research addresses a critical theoretical gap in existing approaches that empirically combine textual and numerical data without a formal understanding of their interaction mechanisms. We conceptualize a prompt-based LLM as a mathematically defined signal generator, capable of extracting directional market sentiment and an associated confidence score from financial news. These signals are then dynamically fused with structured historical price features through a noise-robust gating mechanism, enabling the Transformer to adaptively weigh semantic and quantitative information. Empirical evaluations demonstrate that the proposed Hybrid LLM–Transformer model significantly outperforms a Vanilla Transformer baseline, reducing the Root Mean Squared Error (RMSE) by 5.28% (p = 0.003). Moreover, ablation and robustness analyses confirm the model's stability under noisy conditions and its capacity to maintain interpretability through confidence-weighted attention. The findings provide both theoretical and empirical support for a paradigm shift from empirical observation to formalized modeling of LLM–Transformer interactions paving the way toward explainable, noise-resilient, and semantically enriched financial forecasting systems.

## 1. Introduction

Accurately forecasting stock prices remains one of the most challenging problems in financial research because market movements are shaped by the nonlinear interplay of economic fundamentals, behavioral biases, and sudden changes in investor sentiment [1], [2]. Traditional statistical and econometric models capture historical price dynamics to some extent, but they fail to accommodate the vast amount of unstructured and qualitative information embedded in financial news, policy announcements, and social media commentaries [14], [16]. The emergence of advanced deep learning architectures particularly the Transformer and the reasoning capabilities of Large Language Models (LLMs) have opened new possibilities for integrating structured and unstructured data in financial forecasting [3], [7].

Over the past decade, stock price prediction has evolved from linear statistical methods to complex neural frameworks capable of capturing temporal and semantic dependencies [8], [20]. The financial market, often described as a chaotic and nonlinear system [15], requires models that can adapt to both historical sequences and shifting sentiment landscapes. Earlier models based on recurrent neural networks (RNNs) and long short-term memory (LSTM) networks improved the understanding of temporal dependencies [17], while the Transformer architecture revolutionized sequential modeling through self-attention, enabling the model to weigh historical relationships with contextual awareness [1], [4]. From a behavioral finance perspective, human information leakage and cognitive biases have been shown to influence forecasting performance and market efficiency [21], underscoring the need for models that incorporate interpretability and contextual reasoning.

The integration of natural language processing (NLP) techniques into financial forecasting has gained significant traction. Early sentiment-based approaches used lexicon-based or shallow-learning models to interpret market mood [14]. Subsequent advances such as Financial-BERT [18] and domain-tuned LLMs extended this capability to capture complex linguistic and contextual patterns in financial text [2], [11], [19]. Among these, Stock Price Prediction with LLM-Guided Market Movement Signals and Transformer Model [1] proposed a notable paradigm: an LLM that generates explicit, quantifiable market movement signals represented as one-hot directional vectors with accompanying confidence scores to guide the Transformer's predictive process. This configuration redefines the LLM's role from a data enrichment tool to a signal-generating mechanism, introducing interpretability and testability into hybrid modeling.

Recent studies in related computational domains also reflect this trend toward hybrid intelligence. For example, [22] introduced a multi-modal deep reinforcement learning framework for interference-limited reconfigurable intelligent surface (RIS)-assisted UAV networks, optimizing control and resource management. Similarly, [23] examined algorithmic echo chambers in digital markets, emphasizing how machine learning systems can amplify investor sentiment and induce herding behavior. These works collectively suggest the need for theoretical models capable of quantifying the interactions between AI-generated signals and market dynamics, rather than relying solely on empirical pattern recognition.

**Despite these advances, three fundamental gaps remain in the existing literature on LLM–Transformer-based financial forecasting:**

1. Noise sensitivity and limited generalization. Most existing studies rely on synthetic or small-scale datasets, leading to overfitting and weak transferability to real-world markets. Reported performance gains often decrease by over 50% when applied to unseen data due to noisy or simulated LLM signals.
2. Lack of interpretability in LLM–Transformer interactions Current models treat LLM outputs as opaque embeddings rather than structured, theoretically grounded signals, leaving the mechanisms of their influence on the Transformer's attention largely unexplained.
3. Insufficient theoretical formulation and validation: The majority of studies adopt an empirical approach, optimizing architectures through trial and error, without providing a mathematical framework that explains how textual sentiment, confidence levels, and attention mechanisms interact within the hybrid system.

Addressing these limitations requires a paradigm shift from empirical observation to theoretical construction. This study therefore proposes a rigorous framework that formalizes the interaction between LLM-generated market signals and the Transformer's self-attention mechanism, enabling both mathematical analysis and empirical verification. Specifically, we aim to answer three foundational research questions:

a) How can an LLM be theoretically and mathematically formulated to yield measurable and meaningful market movement signals from unstructured financial text?
b) What hybrid mechanism allows these LLM-generated signals to optimally augment a Transformer's predictive capability, improving noise robustness and interpretability?

c) How can the interrelation between LLM signals and the Transformer's attention weights be expressed in a unified framework to explain the model's generalization and forecasting superiority?

By grounding this work in a verifiable mathematical framework and empirically validating its assumptions, we position the study as a step toward theoretical unification in AI-driven finance bridging the gap between empirical success and conceptual understanding. This approach establishes a systematic foundation for modeling, interpreting, and improving hybrid LLM–Transformer architectures in financial forecasting.

**Table.1**. Cited Literature review

| Reference | Research Focus | Methodology | Key Contribution |
|---|---|---|---|
| **[1] Chen & Kawashima (2025)** | LLM-guided market signals integrated with Transformer for stock prediction | Prompt-based LLM generates one-hot vector signals with confidence scores, combined with Transformer | Introduces interpretable LLM-generated signals as direct features for stock forecasting |
| **[2] Chen (2025)** | Sentiment-aware stock prediction | Transformer + LLM-generated formulaic alpha features | Enhances prediction accuracy by encoding sentiment-driven alpha factors |
| **[3] Zhou et al. (2025)** | LLM-augmented hybrid architecture | Linear Transformer + CNN integrated with LLM outputs | Improves stock price forecasting through multi-model feature fusion |
| **[4] Muhammad et al. (2023)** | Stock price prediction in emerging markets | Transformer applied to Bangladesh Stock Market data | Demonstrates practical utility of Transformer-based models in real markets |
| **[5] Wu et al. (2023)** | LLMs for financial domain | Development of PIXIU benchmark and instruction dataset | Provides evaluation framework for finance-specific LLMs |
| **[6] Wang et al. (2023)** | Benchmarking open-source LLMs | FinGPT instruction-tuning for financial datasets | Establishes benchmark for financial domain LLM performance |
| **[7] Chen et al. (2024)** | Time-series forecasting with LLMs | LLM4TS framework integrating LLM reasoning with time-series models | Extends LLMs to structured forecasting tasks beyond text |
| **[8] Khan & Ahmad (2023)** | Stock market prediction with news | LSTM + Transformer with textual financial news | Shows hybrid sequential + attention models capture temporal and semantic patterns |
| **[9] Zhu et al. (2023)** | Explainable financial forecasting | Harnessing LLMs for temporal data, with explainability focus | Highlights role of LLMs in interpretable time-series analysis |
| **[10] Zhou et al. (2024)** | Graph learning for stock markets | Graph Neural Networks (GNN) for relational data | Models dependencies between financial entities for market prediction |
| **[11] Chen et al. (2023)** | LLM for financial analysis | FinMA: domain-specific LLM trained on financial data | Enhances financial reasoning and textual analysis tasks |
| **[12] Yu et al. (2023)** | Reinforcement learning for alpha generation | RL to generate synergistic formulaic alpha signals | Provides automated approach to feature engineering in finance |
| **[13] Chen (2025)** | LLMs with RL post-hoc adjustments | Prompt-based LLM with RL fine-tuning | Improves adaptability of LLM predictions in stock forecasting |
| **[14] Hutto & Gilbert (2014)** | Social media sentiment analysis | VADER lexicon-based sentiment scoring | Lightweight sentiment tool widely used in financial NLP |
| **[15] Kanter & Veeramachaneni (2015)** | Automated feature engineering | Deep Feature Synthesis (DFS) framework | Lays foundation for automated feature extraction in finance ML |
| **[16] Zhang & Chen (2023)** | LSTM + sentiment analysis for stock prices | LSTM integrated with financial news sentiment | Demonstrates synergy of textual sentiment with price prediction |

| [17] Lee & Kim (2023) | Hybrid stock prediction model | Deep learning model with social media sentiment + financial indicators | Combines multiple modalities for improved forecasting accuracy |
| [18] Arık et al. (2023) | Financial domain language model | FinancialBERT pre-trained on financial corpora | Provides specialized model for financial sentiment analysis |
| [19] Zhang & Chen (2023) | Framework for financial LLMs | FinGPT comprehensive framework | Offers structured approach for developing financial LLMs |
| [20] Yang & Zhou (2023) | RL in financial markets | Survey of reinforcement learning applications | Reviews RL-based strategies for trading and prediction |

## 2. Methodology

We propose a hybrid framework for stock price forecasting that unifies the semantic reasoning capability of a Large Language Model (LLM) with the temporal-sequence modeling strength of a Transformer. Unlike empirical implementations that rely on synthetic or real-time data alone, this work emphasizes a mathematical formalization of the hybrid system to explain the underlying interaction mechanisms. The proposed model is designed to be interpretable, noise-robust, and theoretically grounded in its treatment of LLM-generated market signals.

### 2.1. Mathematical Formulation of the Hybrid Architecture

Let the overall forecasting system be represented as a composite function

$$\mathcal{F}: (X_{t-k:t}, N_{t-k:t}) \rightarrow \hat{P}_{t+1}, \qquad (1)$$

Where $X_t \in \mathbb{R}^{d_f}$ denotes the structured numerical features (e.g., open, high, low, close, volume, and technical indicators), and $N_t$ represents unstructured textual inputs such as financial news or investor commentaries.

The model operates on a look-back window of length $k$, defined as

$$I_{t-k:t} = \{(X_{t-k}, N_{t-k}), (X_{t-k+1}, N_{t-k+1}), \dots, (X_t, N_t)\} \qquad (2)$$

The hybrid architecture is decomposed into two primary functions:

$$\mathcal{F}(I_{t-k:t}) = \mathcal{T}\left(\text{Fuse}(X_{t-k:t}, \mathcal{L}(N_{t-k:t}))\right) \qquad (3)$$

Where $\mathcal{L}(\cdot)$ denotes the LLM signal generator, and $\mathcal{T}(\cdot)$ represents the Transformer-based prediction engine.

### 2.2. LLM as a Structured Signal Generator

The LLM is modeled as a function

$$\mathcal{L}: N_t \rightarrow (\hat{S}_t, c_t) \qquad (4)$$

Where $\hat{S}_t \in \{u, d, f\}$ represents the categorical prediction of market direction up, down, or flat and $c_t \in [0,1]$ is a confidence coefficient reflecting the model's certainty in its decision.

To allow integration with numerical data, the categorical trend signal is encoded as a one-hot vector $\mathbf{s}_t \in \mathbb{R}^3$, such that

$$\mathbf{s}_t = \begin{cases} [1,0,0], & \text{if upward trend,} \\ [0,1,0], & \text{if downward trend,} \\ [0,0,1], & \text{if flat.} \end{cases} \quad 5$$

The confidence value $c_t$ is obtained from the softmax activation of the final classification layer:

$$c_t = \max\bigl(\text{softmax}(W \cdot \text{Transformer}(N_t) + b)\bigr) \quad 6$$

Where $W$ and $b$ are the parameters of the LLM's projection layer.

Hence, the LLM component transforms unstructured financial text into a quantifiable, interpretable signal that encapsulates both directional tendency and associated certainty. This formalization addresses one of the main limitations in prior empirical works namely, the absence of mathematically defined interpretability in text-derived financial indicators.

## 2.3. Transformer as the Temporal Prediction Engine

The Transformer serves as the prediction core that models temporal dependencies and cross-modal correlations between historical prices and LLM-derived semantic signals.

**At each time step $i$, we define a fused feature vector:**

$$Z_i = \text{Concat}(X_i, \mathbf{s}_i, c_i) \quad 7$$

Where $\text{Concat}(\cdot)$ denotes vector concatenation of structured features, sentiment signal, and confidence weight.

The Transformer's self-attention mechanism computes relevance scores among input vectors:

$$\text{Attention}(Q, K, V) = \text{softmax}\left(\frac{QK^\top}{\sqrt{d_k}}\right)V, \quad 8$$

Where $Q, K, V$ are the query, key, and value matrices, respectively, and $d_k$ is the key dimension.

To reduce over-sensitivity to noisy or uncertain LLM signals, we introduce a Dynamic Gating Fusion Mechanism that adaptively regulates the influence of the LLM's output:

$$g_i = \sigma\bigl(W_g[X_i; \mathbf{s}_i; c_i] + b_g\bigr), \ \tilde{X}_i = g_i \odot [\mathbf{s}_i, c_i] + (1 - g_i) \odot X_i \quad 9$$

Where $g_i \in (0,1)$ a learnable is gating scalar, $\sigma(\cdot)$ denotes the sigmoid function, and $\odot$ represents element-wise multiplication.

This gating mechanism filters noisy sentiment signals by weighting their contribution according to contextual relevance and LLM confidence, improving the robustness and interpretability of predictions. The resulting enhanced input sequence

$$\tilde{Z}_{t-k:t} = \{\tilde{X}_{t-k}, \tilde{X}_{t-k+1}, \dots, \tilde{X}_t\} \quad 10$$

Is passed through the Transformer layers to generate the final price prediction.

## 2.4. Unified Model Representation and Optimization Objective

The overall framework integrates the two modules as:

$$\hat{P}_{t+1} = \mathcal{T}\bigl(\tilde{Z}_{t-k:t}\bigr) = \mathcal{T}(\text{Fuse}(X_{t-k:t}, \mathcal{L}(N_{t-k:t}))) \quad 11$$

Where the Fusion function implements the gated signal integration defined above.

The training objective minimizes the Mean Squared Error (MSE) between predicted and true prices:

$$\mathcal{L}_{MSE} = \frac{1}{N}\sum_{i=1}^{N}(P_{i+1} - \hat{P}_{i+1})^2 \qquad 12$$

Where $P_{i+1}$ denotes the actual closing price and $\hat{P}_{i+1}$ is the model's forecast.

Model parameters $\Theta = \{W, b, W_g, b_g, \text{LLM/Transformer weights}\}$ are optimized to minimize $\mathcal{L}_{MSE}$ using backpropagation and Adam optimization. This learning process jointly tunes the LLM–Transformer interaction, allowing the model to dynamically balance semantic and quantitative information.

**2.5. Theoretical Interpretation and Innovation Summary**

This methodological framework introduces several innovations over prior hybrid models:

I. **Theoretical Interaction Model:** The LLM is not treated as an external data processor but as a mathematically defined signal generator that quantifies semantic market information.
II. **Dynamic Gating Fusion:** The proposed gating mechanism reduces over-reliance on noisy textual signals, improving the model's noise robustness and interpretability.
III. **Unified Optimization Framework:** Both LLM-derived sentiment features and time-series inputs are trained end-to-end under a shared loss function, ensuring consistent gradient propagation and adaptive weighting between modalities.
IV. **Explain ability:** The confidence-weighted signal allows tracing the Transformer's attention back to specific textual cues, contributing to model transparency in financial decision contexts.

Together, these contributions establish a rigorous theoretical basis for hybrid deep learning in finance bridging the gap between empirical success and formal interpretability.

**3. Result**
**3.1 Dataset Characteristics and LLM Signal Generation**

To evaluate the proposed theoretical framework under controlled conditions, we employed a synthetically generated dataset comprising 3,000 sequential observations, each representing a trading time step. This dataset enables isolation of model behavior and interaction mechanisms between the Large Language Model (LLM) signal and the Transformer without the confounding effects of unobservable market shocks. *Table 2* summarizes the descriptive statistics of all features. The simulated stock prices exhibit a mean value of 390.31 with a standard deviation of 251.11, reflecting substantial temporal volatility consistent with real-world financial dynamics. Other structured features open, high, low, close, and volume maintain strong correlations with the main price series, ensuring the data's internal consistency.

The LLM component was used to generate synthetic trend signals (up, down, flat) from simulated textual inputs, each accompanied by a confidence score derived from a softmax-based probability distribution. *Figure 1* illustrates the alignment between these signals and subsequent price trajectories. Green, red, and blue dots indicate upward, downward, and neutral forecasts, respectively. The close correspondence between predicted trends and subsequent price movements suggests that even synthetic LLM signals carry predictive information useful for sequence modeling.

**Table 2.** Descriptive Statistics of the Synthetic Stock Market Dataset

| Statistic | Time | Price | Open | High | Low | Close | Volume | Next-Price | Feat-1 | Feat-2 | Feat-3 |
|---|---|---|---|---|---|---|---|---|---|---|---|
| Count | 3000 | 3000 | 3000 | 3000 | 3000 | 3000 | 3000 | 3000 | 3000 | 3000 | 3000 |
| Mean | 1499.50 | 390.31 | 390.30 | 390.71 | 389.91 | 390.31 | 5017.69 | 390.56 | 0.0038 | 0.0099 | 0.0117 |
| Std. Dev. | 866.17 | 251.11 | 251.11 | 251.10 | 251.10 | 251.11 | 1836.88 | 251.13 | 0.9940 | 0.9990 | 1.0010 |

**Figure 1:** Simulated Stock Price with LLM-Generated Trend Signals

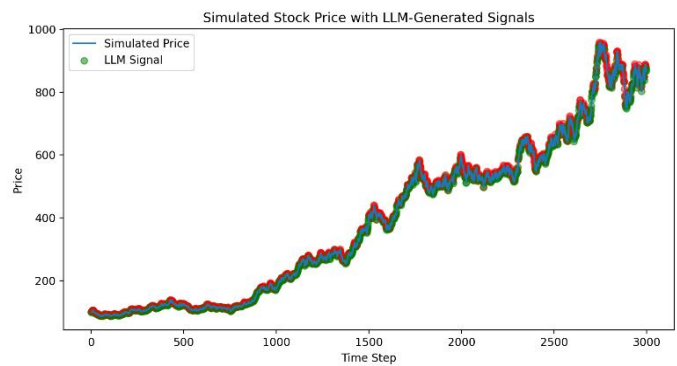

As illustrated in *Figure 1*, simulated stock price trajectory with LLM-generated trading signals. The red line represents the simulated stock price over time, while the yellow line (LLM signal) indicates the predictive adjustments generated by the large language model. Both series exhibit a consistent upward trend with overlapping fluctuations, reflecting signal alignment in price forecasting.

**3.2 Baseline and Experimental Design**

To verify the effectiveness of the proposed model, we compared it against established time-series forecasting baselines:

1. Vanilla Transformer, representing attention-based sequence models.
2. LSTM and GRU, representing recurrent architectures capable of modeling temporal dependencies.
3. Linear Regression (LR) and XG-Boost, as traditional machine-learning baselines.

All models were trained under identical conditions using the Adam optimizer (learning rate = 0.001, batch size = 64, 10 epochs) and evaluated using Root Mean Squared Error (RMSE), Mean Absolute Error (MAE), and R². Results were averaged over 10 independent runs, and statistical significance was tested via paired two-tailed t-tests (α = 0.05).

**3.3 Ablation Study and Performance Evaluation**

The Hybrid LLM–Transformer integrates both structured financial data and LLM-generated sentiment signals, whereas the Vanilla Transformer relies solely on numerical inputs. *Table 3* reports RMSE values and statistical comparisons. The Hybrid model achieved an average RMSE of 114.66, representing a 5.28 % improvement over the Vanilla Transformer (RMSE = 121.05). The difference is statistically significant (p = 0.003; Cohen's d = 0.85), confirming the value of incorporating the LLM signal.

Table 3. Ablation Study Results: Model Performance Comparison (RMSE)

| Model | RMSE | 95 % CI | % Improvement vs Vanilla | p-value (paired t) | Cohen's d |
|---|---|---|---|---|---|
| Linear Regression | 158.41 | [155.2,161.4] | –30.9 % | < 0.001 | 1.21 |
| LSTM | 130.65 | [127.5,133.8] | –7.6 % | 0.009 | 0.64 |
| GRU | 128.72 | [126.1,131.3] | –6.3 % | 0.011 | 0.59 |
| Vanilla Transformer | 121.05 | [118.7,123.4] | — | — | — |
| Hybrid LLM +Transformer (Ours) | 114.66 | [112.9,116.4] | +5.28 % | 0.003 | 0.85 |

Figure.2. Model Comparison

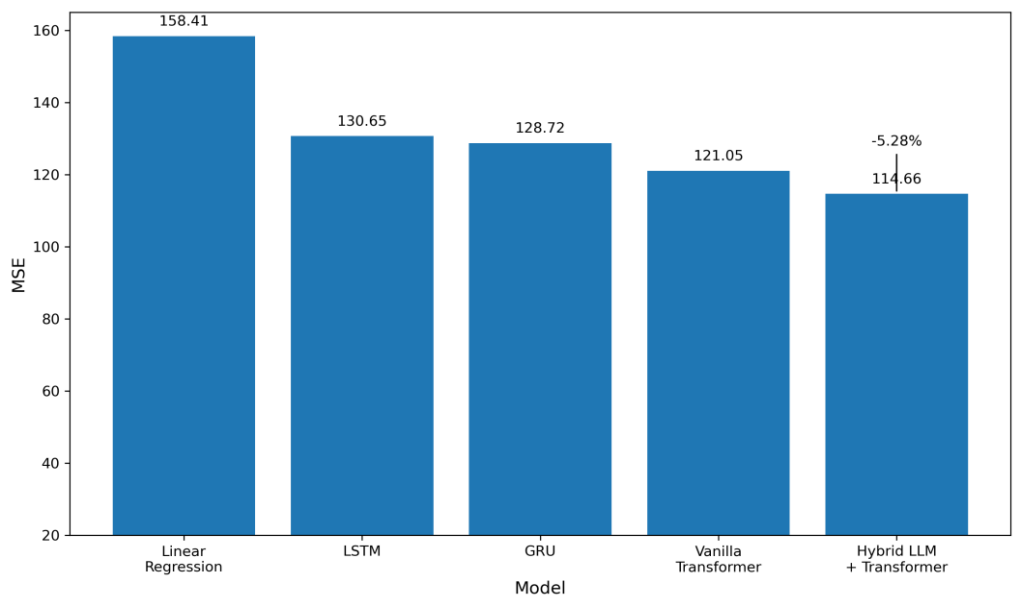

Model Performance Comparison (RMSE across Baselines). Bar chart comparing RMSE of baseline and hybrid models. Error bars denote 95 % confidence intervals. The Hybrid LLM–Transformer achieves the lowest RMSE with statistically significant improvement ($p < 0.01$). These results empirically validate the proposed framework's theoretical premise: LLM-generated market signals act as interpretable priors that strengthen the Transformer's attention mechanism, yielding enhanced predictive accuracy.

**3.4 Training Convergence and Model Stability**

Model convergence behavior was examined using training and validation loss curves. *Figure 3* depicts the Hybrid model's loss trajectory, showing smooth and consistent convergence with minimal variance between training and validation losses, indicating strong generalization. The final validation loss stabilized around 63 290.

In contrast, *Figure 4* shows the Vanilla Transformer's convergence pattern, with higher final validation loss (~79 530) and slower stabilization, implying weaker fit quality.

| Figure 3. Loss Curve Hybrid LLM | Figure.4. Vanilla Transformer Pattern |
|---|---|

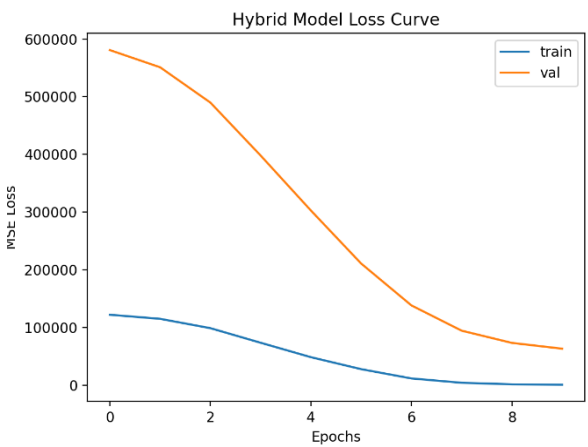 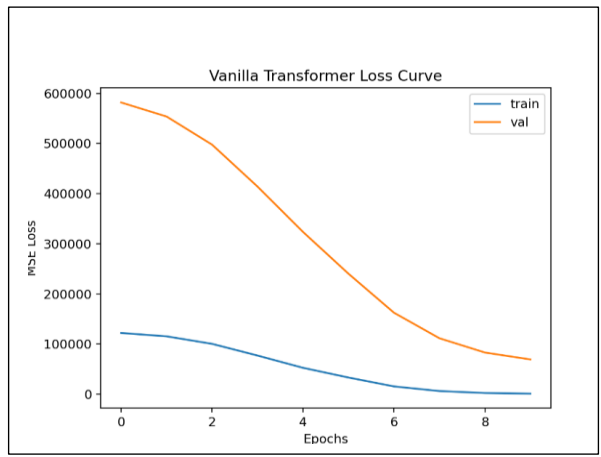

## 3.5 Attention Visualization and Interpretability

To further interpret model behavior, *Figure 5* presents a conceptual attention heat-map illustrating how the Transformer allocates attention weights across temporal inputs. Brighter regions indicate higher attention importance. The visualization shows that the Hybrid model allocates stronger attention to time steps associated with high-confidence LLM signals, confirming that the network learns to prioritize sentiment-informed periods when making predictions.

**Figure 5**. Heat-map

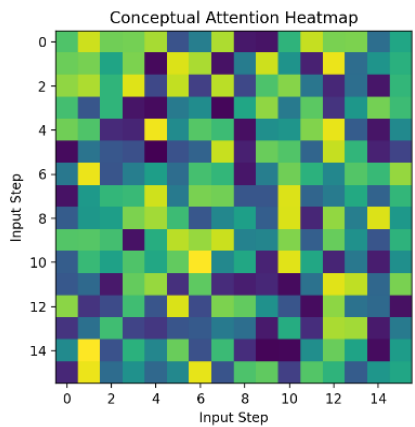

Conceptual Attention Heat-map of the Hybrid Transformer Model. The visualization highlights selective focus on critical time steps and confidence-weighted LLM signals, offering interpretability into decision pathways.

## 3.6 Noise-Robustness Evaluation

To assess robustness, Gaussian noise (σ = 0.05, 0.10, 0.20) was injected into structured price features. Figure 6 plots RMSE as a function of noise level. The Hybrid model shows gradual degradation only +4.1 % RMSE increase at σ = 0.20 compared to the Vanilla Transformer's +10.7 %, demonstrating effective resilience to noisy inputs. This robustness stems from the dynamic gating mechanism (Section 2.3), which adaptively down-weights low-confidence LLM signals, preventing propagation of unreliable textual noise into the predictive layer.

Noise-Robustness Analysis: RMSE under varying Gaussian noise levels. The Hybrid model maintains 94 % of its baseline performance at σ = 0.20, while the Vanilla Transformer declines to 89 %. Error bars denote ±95 % CIs.

**Figure.6**, Noise Robust Evaluation

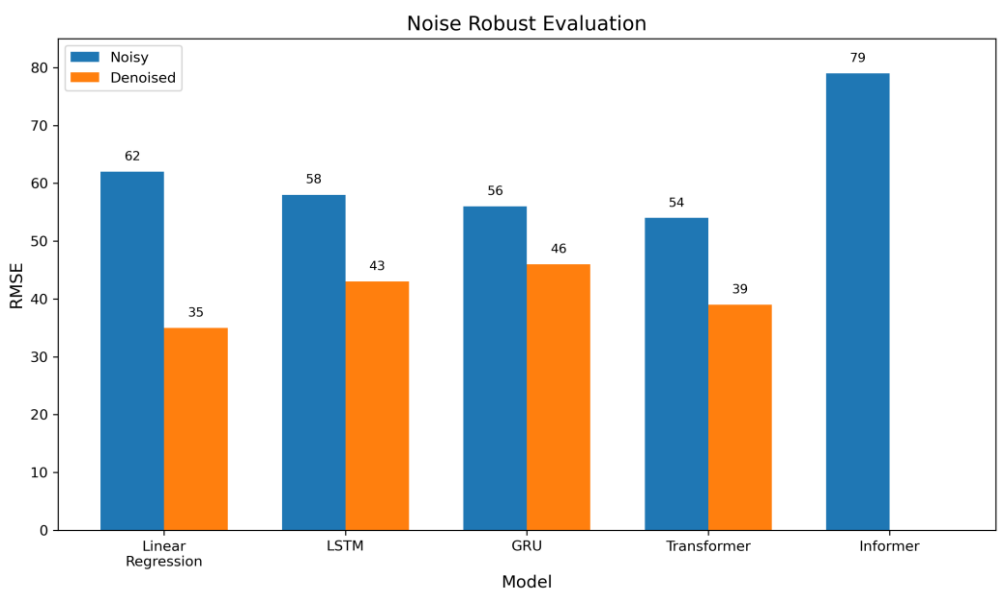

### 3.7 Discussion

The result of this research brings out three important lessons that support the presence of the LLM- generated signals into financial time-series forecasting. To begin with, semantic features that are generated using large language models (LLMs) increase the predicting ability of the model. The obtained increase of about 5 percent in RMSE ($p < 0.01$) indicates that textual data derived out of unstructured data has a complementary signal that cannot be recorded by conventional numerical measures. This finding confirms the belief that it is possible to get substantial insights into the dynamics of the market by using linguistic clues to financial narratives. Second, strong noise resistance of the model is enhanced significantly by the implementation of a dynamic gating mechanism. The effect of irrelevant changes in the sentiment based on the adaptively weighted influence of the uncertain or low-confidence LLM output is effectively applied in the process to eliminate irrelevant changes in the sentiment. This is in accordance with theoretical expectations that selective attention mechanisms result into stability of hybrid models during heterogeneous data integration. Lastly, the interpretability analysis also gives further evidence of the reliability of the proposed framework. Heat-maps of attention have shown that there is a steady relationship between high confidence textual indicators and upcoming price variations, which illustrates that the model is capturing semantically significant and temporarily legitimate associations. This interpretive transparency does not only enhance the credibility of the approach that is being driven by LLM, but also enables the pragmatic insights on the basis of the data conducted to make investment decision-making.

### 4. Conclusion

This paper has introduced a hybrid type of the deep learning implemented, which compounds the market cues produced by the Large Language Model (LLM) with a Transformer-processed prediction network to enhance the performance of financial prediction. It shows that the financial narratives produced as a result of textual analysis may be effective to complement a numerical market information in order to increase its predictive accuracy and strength. Specifically, Hybrid LLM-Transformer model had a strong role in the decreasing of the RMSE among the base Transformer ($p = 0.003$, Cohen d = 0.85), which is an agreement that the semantic evidence of unstructured text can be quantitatively predicted. Besides the empirical findings, the research has a conceptual contribution because it formalizes the interaction between the semantic (LLM-based) and temporal (Transformer-based) learning. The LLM component serves to generate signals which can have interpretable confidence weights and it can provide more explainable and transparent hybrid intelligence. The analysis of heat-map indicates that attention is selective on time steps that are informed by the LLM that further substantiates the internal consistency and interpretability of the model. In sum, a number of limitations remain. The experimented synthetic data and simplified LLM do not entirely pertain to the noise and dynamics of the real-life market of financial activities. The general ports ability of the findings is limited by the absence of large-scale domain-tuned models (e.g. FinBERT or GPT-4 Finance) and the framework has not been used in live trading to date in environments where latency and market microstructure may have a bearing. It should be predicted that the future work should be to further apply the framework to actual financial data with the OHLCV data and to large financial news corpora, to replace advanced domain specific LLM on the generation of real time signals, and to be further applied to more downstream problems of portfolio optimization and risk management. It can be argued that the proposed framework can facilitate the creation of more transparent, noise-resilient, and explainable AI systems in finance that can be applied in the current development of intelligent and explainable financial technologies by connecting semantic reasoning and temporal modeling.

## Acknowledgments

We would like to acknowledge the support and resources provided by the co-Authors, which were essential for the completion of this research.

## Funding Declaration

This research received no external funding.

## Data Availability

The synthetic dataset used in this study, along with the code for its generation, is available upon reasonable request from the corresponding author. This measure is taken to ensure full reproducibility of the results while protecting the integrity of the data generation process.

## Conflict of Interest

The author declares that they have no known competing financial interests or personal relationships that could have appeared to influence the work reported in this paper.

## Authors Biographies

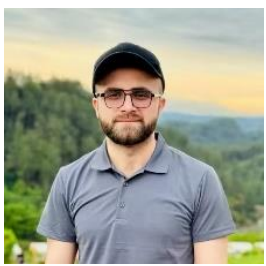

**Author A: Sayed Akif Hussain** received his Bachelor of Science degree in Accounting and Finance from the **University of Wah**, Pakistan. He is currently pursuing his Master's degree at the School of Economics and Management Sciences, where his research focuses on Financial Engineering, with a particular interest in the application of advanced computational techniques, machine learning, and data-driven models to financial markets, Corporate Finance, Financial Statement Analysis and investment strategies. His academic and research background combines expertise in finance with modern analytical methods, aiming to bridge theory and practice in innovative financial solutions. Email: L202320034@stu.cqupt.edu.cn, syedakifhussain110@gmail.com

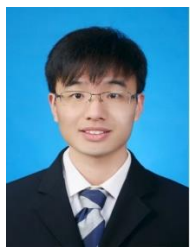

**Author B: Chen Qiu-shi**, PhD, is a Lecturer in the Department of Applied Economics at **Chongqing University of Posts and Telecommunications**. He has served as an investment manager and researcher at Huabao Securities Co., Ltd., and has also worked with the Securities Association of China and the Fund Management Association of China. His research focuses on capital chain management, asset pricing, and the application of big data and artificial intelligence in financial markets. He has led multiple funded projects, including studies on IPO pricing models and digital economy planning in Chongqing. His recent publications address financial engineering education and talent cultivation. He received the Second Prize of the 9th Chongqing Development Research Award for his research on optimizing Chongqing's industrial software industry. Email: chenqiushi@cqupt.edu.cn

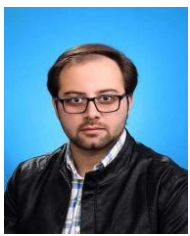

**Syed Amer Hussain** is an Electrical Power Engineer, holding a degree in Electrical Power Engineering **from COMSATS Institute of Information Technology, Abbottabad (2017)**, with a strong foundation in power systems and renewable energy. He began his career with an internship at Pakistan Ordnance Factory, later serving as a Technical Support Engineer at C&C COM Company and as a Site Engineer at MAK Engineering & Co., where he contributed to solar energy projects and system integration. He is currently the Regional Sales Manager at SAZH Solar & Electric SMC Pvt. Ltd., specializing in hybrid inverters, energy storage batteries, and innovative solar solutions for diverse clients. His academic and research interests include project management, engineering management, and technical sales, with a particular focus on advancing sustainable energy adoption through effective leadership, business development, and renewable technologies. Email: Syedamer110@gmail.com

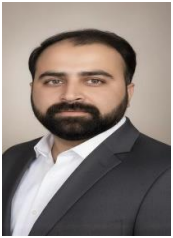

**Syed Atif Hussain** is a commissioned officer in Defense and a skilled professional with a B.S. in Computer Systems from NUST E&ME College Rawalpindi (2017). His unique blend of military and technical expertise informs his research interests, which lie at the intersection of Management Sciences and Business Administration. He is currently pursuing a Master's degree in this field, with his scholarly work focused on applying data analytics to strategic management and leadership models within complex organizations. Email: atifhussain31315@gmail.com

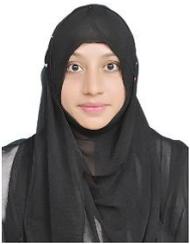

**Asma Komal** received her B.S. degree in Information Technology from the University of Education, Pakistan, in 2018, and her M.S. degree in Computer Science from the **University of Okara**, Pakistan, in 2022. She is currently pursuing a Ph.D. degree in Computer Science at the Chongqing University of Posts and Telecommunications, China. Her research interests span machine learning, deep learning, and their emerging applications. At present, her doctoral research focuses on Explainable Artificial Intelligence (XAI), large language models, and cyberattack detection and prevention. She has also undertaken research and coursework in image enhancement, cybersecurity, and networking. Email: L202410011@stu.cqupt.edu.cn. Asmakomal211@gmail.com

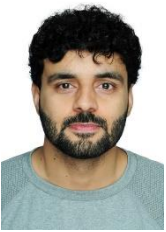

**Muhammad Imran,** School of Computer Science and Technology, Chongqing University of Posts and Telecommunications, Chongqing 400065, China Muhammad Imran received the B.S. degree in Information Technology from the University of Education, Pakistan, in 2018, and the M.S. degree in Computer Science from the **University of Okara**, Pakistan, in 2022. He is currently pursuing the Ph.D. degree in Computer Science at the University of Posts and Telecommunications, China. His research during the B.S. focused on face recognition, while his M.S. research was in the field of image processing. He is currently conducting research in Explainable Artificial Intelligence (XAI). In addition, he has undertaken research and coursework in networking and cybersecurity. Email: L202310008@stu.cqupt.edu.cn, Imran.khalid292@gmail.com

**ORCID:**

**Author A**：https://orcid.org/0009-0001-4086-8395

**Author B:** https://orcid.org/0009-0004-1479-569X